\documentclass{llncs}

\usepackage{amsmath}
\usepackage{graphicx}
\usepackage{xspace}
\usepackage{cite}
\usepackage[utf8]{inputenc}
\usepackage[T1]{fontenc}
\usepackage{url}

\newcommand\tstrut{\rule{0pt}{2.4ex}}

\makeatletter
\spn@wtheorem{observation}{Observation}{\bfseries}{\itshape}
\makeatother

\title{Efficient Representation for \penalty-1 Online Suffix Tree Construction}
\author{N.\,Jesper Larsson \and Kasper Fuglsang \and Kenneth Karlsson}

\institute{IT University of Copenhagen, Denmark,\\
\email{\{jesl,kfug,kkar\}@itu.dk}}

\newcommand\emptystring{{\epsilon}}
\newcommand\ST{{\mathcal{ST}}}
\newcommand\STi[1]{{\mathcal{ST}_{\kern-.222em #1}}}

\newcommand\alphabet{{\mathrm{\Sigma}}}
\newcommand\reprupd{{\mathbf{repr\kern-.1em\textit{-}update}}}
\newcommand\mrmfind{{\mathbf{mrm\kern-.1em\textit{-}find}}}

\let\bibtla\textsc

\newcommand\notd{\textsc{notd}\xspace}
\newcommand\nobu{\textsc{nobu}\xspace}
\newcommand\eotd{\textsc{eotd}\xspace}
\newcommand\eov{\textsc{eov}\xspace}
\newcommand\lili{\textsc{ll}\xspace}
\newcommand\hata{\textsc{ht}\xspace}
\newcommand\cpu{\textsc{cpu}\xspace}

\DeclareMathSymbol{\topnode}{\mathord}{symbols}{"3F}
\DeclareMathSymbol{\topedge}{\mathord}{symbols}{"60}

\begin{document}

\maketitle

\begin{abstract}
  Suffix tree construction algorithms based on \emph{suffix links} are popular because
  they are simple to implement, can operate \emph{online} in linear time, and because the suffix links are
  often convenient for pattern matching. We present an approach using
  \emph{edge-oriented} suffix links, which reduces the number of branch
  lookup operations (known to be a bottleneck in construction time) with some
  additional techniques to reduce construction cost. We discuss various effects of our approach and
  compare it to previous techniques.  An experimental evaluation shows that we
  are able to reduce construction time to around half that of the original algorithm, and
  about two thirds that of previously known branch-reduced construction.
\end{abstract}

\section{Introduction}\label{sec-intro}

The \emph{suffix tree} is arguably the most important data structure in string
processing, with a wide variety of
applications~\cite{Apostolico85,gusfield,SufComp}, and with a number of
available construction
algorithms~\cite{Weiner73,McR,UkkoOnli,FarFOCS,giegkurtzstoyetopdown,canovas2010practical},
each with its benefits. Improvements in its efficiency of construction and
representation continues to be a lively area of research, despite the fact that
from a classical asymptotic time complexity perspective, optimal solutions have
been known for decades. Pushing the edge of efficiency is critical for indexing
large inputs, and make large amounts of experiments feasible, e.g., in
genetics, where lengths of available genomes increase. Much work has been
dedicated to reducing the memory footprint with representations that are
compact~\cite{kurtzsuftree} or compressed (see C{\'a}novas and
Navarro~\cite{canovas2010practical} for a practical view, with references to
theoretical work), and to alternatives requiring less space, such as suffix
arrays~\cite{Manber93}.  Other work adresses the growing performance-gap
between cache and main memory, frequently using algorithms originally designed
for secondary
storage~\cite{ClarkMunro,ferraginaHierarchxs,TsirogiannisModernSuffix,TianSuffixVLDB}.

While memory-reduction is important, it typically requires elaborate operations
to access individual fields, with time overhead that can be deterring for some
applications. Furthermore, compaction by a reduced number of pointers per node
is ineffective in applications that use those pointers for pattern matching. Our
work ties in with the more direct approach to improving performance of the
conventional primary storage suffix tree representation, taken by Senft and
Dvořák~\cite{SenftBranching}. Classical representations required in Ukkonen's
algorithm~\cite{UkkoOnli} and the closely related predecessor of
McCreight~\cite{McR} remain important in application areas such as genetics,
data compression and data mining, since they allow online construction as well
as provide \emph{suffix links}, a feature useful not only in construction,
but also for string matching tasks~\cite{gusfield,kielbasa2011adaptive}. In
these algorithms, a critically time-consuming operation is \emph{branch}:
identifying the correct outgoing edge of a given node for a given
character~\cite{SenftBranching}. This work introduces and evaluates several
representation techniques to help reduce both the number of branch operations
and the cost of each such operation, focusing on running time, and taking an
overall liberal view on space usage.

Our experimental evaluation of runtime, memory locality, and the counts for
critical operations, shows that a well chosen combination of our presented
techniques consistently produce a significant advantage over the original
Ukkonen scheme as well as the branch-reduction technique of Senft and Dvořák.

\section{Suffix Trees and Ukkonen's Algorithm}\label{sec-defs}

We denote the \emph{suffix tree} (illustrated in fig.~\ref{fig-st}) over a
string $T=t_0\cdots t_{N-1}$ of length $|T|=N$ by $\ST$\!. Each edge in
$\ST$\!, directed downwards from the root, is labeled with a substring of
$T$\!, represented in constant space by reference to position and length in
$T$\!. We define a \emph{point} on an $\ST$ edge as the position between two
characters of its label, or -- when the point coincides with a node -- after
the whole edge label. Each point in the tree corresponds to precisely one
nonempty substring $t_i\cdots t_j$, $0\le i\leq j<N$, obtained by reading edge
labels on the path from the root to that point. A consequence is that the first
character of an edge label uniquely identifies it among the outgoing edges of a
node. The point corresponding to an arbitrary pattern can be located (or found
non-existent) by scanning characters left to right, matching edge labels from
the root down.  For convenience, we add an auxiliary node $\topnode$ above the
root (following Ukkonen), with a single edge to the root. We denote this edge
$\topedge$ and label it with the empty string, which is denoted by
$\emptystring$. (Although $\topnode$ is the topmost node of the augmented tree,
we consistently refer to the root of the unaugmented tree as the root of
$\ST$\!.) Each leaf corresponds to some suffix~$t_i\cdots t_{N-1}$, $0\le
i<N$. Hence, the label endpoint of a leaf edge can be defined implicitly,
rather than updated during construction. Note, however, that any suffix that is
not a unique substring of~$T$ corresponds to a point higher up in the tree. (We
do not, as is otherwise common, require that $t_{N-1}$ is a unique character,
since this clashes with online construction.)

Except for $\topedge$, all edges are labeled with nonempty strings,
and the tree represents exactly the substrings of $T$ in the minimum number of
nodes. This implies that each node is either $\topnode$, the root, a leaf, or a
non-root node with at least two outgoing edges. Since the number of leaves is
at most $N$ (one for each suffix), the total number of nodes cannot exceed
$2N+1$ (with equality for $N=1$).


We generalize the definition to $\STi{i}$ over string $T_i=t_0 \cdots
t_{i-1}$, where $\STi{N}=\ST$. An \emph{online} construction algorithm
constructs $\ST$ in $N$ updates, where update $i$ reshapes
$\STi{i-1}$ into $\STi{i}$, without looking ahead any further than $t_{i-1}$.

We describe suffix tree construction based on Ukkonen's algorithm~\cite{UkkoOnli}. Please refer to Ukkonen's original,
closer to an actual implementation, for details such as
correctness arguments.

 Define the
\emph{active point} before update $i>1$ as the point corresponding to the
longest suffix of $T_{i-1}$ that is not a unique substring of
$T_{i-1}$. Thanks to the implicit label endpoint of leaf edges, this is the point
of the longest string where update $i$ might alter the tree. The active
point is moved once or more in update $i$, to reach the corresponding start
position for update $i+1$. (This diverges
slightly from Ukkonen's use, where the active point is only defined as the
start point of the update.)  Since any leaf corresponds to a
suffix, the label end position of any point coinciding with a leaf in $\STi{i}$
is $i-1$. The tree is augmented with \emph{suffix links},
pointing upwards in the tree: Let $v$ be a non-leaf node that coincides with
the string $aA$ for some character $a$ and string $A$. Then the suffix link of
$v$ points to the node coinciding with the point of $A$. The suffix link of the
root leads to $\topnode$, which has no suffix link. Before the first update,
the tree is initialized to $\STi{0}$ consisting only of $\topnode$ and the
root, joined by $\topedge$, and the active point is set to the endpoint of
$\topedge$ (i.e.\@ the root). Update $i$ then procedes as
follows:

\begin{enumerate}
\item If the active point coincides with $\topnode$, move it down one step to
  the root, and finish the update.
\item Otherwise, attempt to move the active point one step down, by scanning
  over character $t_i$. If the active point is at the end of an edge, this
  requires a \emph{branch} operation, where we choose among the outgoing
  edges of the node. Otherwise,
  simply try matching the character following the point with $t_i$. If the move down
  succeeds, i.e., $t_i$ is present just below the active point, the update
  is finished. Otherwise, keep the current active point for now, and continue with
  the next step.
\item\label{step-ukksplit} Unless the active point is at the end of an edge,
  split the edge at the active point and introduce a new node. If there is a
  saved node $v_p$ (from step~\ref{step-ukksavevp}), let $v_p$'s suffix link
  point to the new node. The active point now coincides with a node, which we
  denote $v$.
\item Create a new leaf $w$ and make it a child of $v$. Set the start pointer
  of the label on the edge from $v$ to $w$ to $i$ (the end pointer of leaf
  labels being implicit).
\item If the active point corresponds to the root, move it to
  $\topnode$. Otherwise, we should move the active point to correspond
  to the string $A$, where $aA$ is the string corresponding to $v$ for some character
  $a$. There are two cases:
  \begin{enumerate}
  \item Unless $v$ was just created, it has a suffix link, which
    we can simply follow to directly arrive at a node that coincides with the
    point we seek.
  \item\label{step-ukkrescan} Otherwise, i.e. if $v$'s suffix link is not yet set, let $u$ be
    the parent of $v$, and follow the suffix link of $u$ to $u'$. Then locate
    the edge below $u'$ containing the point that corresponds to $A$. Set this
    as the active point. Moving down from $u'$ requires one or more branch
    operations, a process referred to as \emph{rescanning} (see fig.~\ref{fig-linkstyles}). If
    the active point now coincides with a node $v'$, set the suffix link of $v$
    to point to $v'$. Otherwise, save $v$ as $v_p$ to have its suffix link set
    to the node created next.\label{step-ukksavevp}
  \end{enumerate}
\item Continue from step~1.
\end{enumerate}

\section{Reduced Branching Schemes}\label{sec-redbranch}

Senft and Dvořák~\cite{SenftBranching} observe that the \emph{branch} operation,
searching for the right outgoing edge of a node, typically dominates execution
time in Ukkonen's algorithm. Reducing the cost of branch can significantly improve
construction efficiency. Two paths are possible: attacking the cost of the
branch operation itself, through the data structures that support it, which
we consider in section~\ref{sec-hashvsll}, and reducing the \emph{number}
of branch operations in step~\ref{step-ukkrescan} of
the update algorithm.

We refer to Ukkonen's original method of maintaining and using suffix links as
\emph{node-oriented top-down} (\notd). Section~\ref{sec-nobu} discusses the
\emph{bottom-up} approach (\nobu) of Senft and Dvořák, and
sections~\ref{sec-eotd}--\ref{sec-eov} present our novel approach of
\emph{edge-oriented} suffix links, in two variants \emph{top-down} (\eotd) and
\emph{variable} (\eov).

\subsection{Node-Oriented Bottom-Up}\label{sec-nobu}

A branch operation comprises the rather expensive task of locating, given a
node $v$ and character $c$, $v$'s outgoing
edge whose edge label begins with $c$, if one exists. By contrast, following an
edge in the opposite direction can be made much cheaper, through a parent pointer. Senft and Dvořák~\cite{SenftBranching}
suggests the following simple modification to suffix tree representation and
construction:
\begin{itemize}
\item Maintain parents of nodes, and suffix links for
  leaves as well as non-leaves.
\item In step~\ref{step-ukkrescan} of
update, follow the suffix link of $v$ to $v'$ rather than that of its
parent $u$ to $u'$, and locate the point corresponding to $A$ moving up,
\emph{climbing} from $v'$
rather than rescanning from $u'$ (see fig.~\ref{fig-linkstyles}).
\end{itemize}

Senft and Dvořák experimentally demonstrate a runtime improvement across a
range of typical inputs. A drawback is that worst case time complexity is
not linear: a class of inputs with time complexity $\Omega(N^{1.5})$ is
easily constructed, and it is unknown whether actual worst case
complexity is even higher. To circumvent degenerate cases, Senft and Dvořák
suggest a hybrid scheme where climbing stops after $c$ steps, for constant $c$, falling back to rescan. (As an alternative, we
suggest bounding the number of edges to climb to by using rescan iff the remaining edge label length below the active point
exceeds constant $c'$.) Some of the space overhead can be avoided in a representation using clever
leaf numbering.

\subsection{Edge-Oriented Top-Down}\label{sec-eotd}

We consider an alternative branch-saving strategy, slightly modifying
suffix links.

For each split edge, the \notd update algorithm follows a suffix link from
$u$ to $u'$, and immediately obtains the outgoing edge $e'$ of
$u'$ whose edge label starts with the same character as the edge just
visited. We can avoid this
\emph{first} branch operation in rescan (which constitutes a large part of
rescan work)
, by having $e'$ available from $e$ directly, without taking the detour via
$u$ and $u'$.

\begin{figure}[t]
  \begin{center}
\includegraphics[scale=.9]{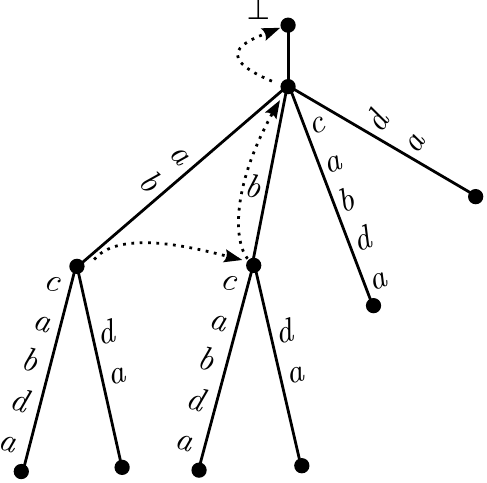}
\hfil
\includegraphics[scale=.9]{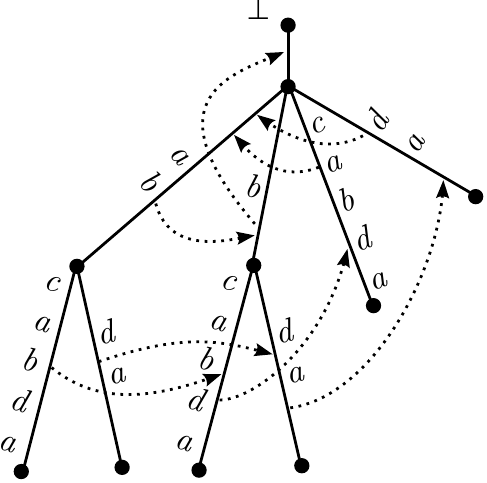}
\end{center}
\caption{\label{fig-st}Suffix tree over the string $abcabda$, with dotted lines
  showing node-oriented suffix links for internal nodes only, as in Ukkonen's
  original scheme (left), and
  edge-oriented suffix links (right).}
\end{figure}

Define the string that \emph{marks} an edge as the shortest string represented
by the edge
(corresponding to the point after one character in its label). For edge $e$,
let $aA$, for character $a$ and string $A$, be the shortest string represented
$e$ such that $A$ marks some other edge $e'$. (The same as saying that $aA$
marks $e$, except when $e$ is an outgoing edge of the root and $|A|=1$, in
which case $a$ marks $e$.) Let the \emph{edge oriented suffix link} of $e$
point to $e'$ (illustrated i fig.~\ref{fig-st}).

\label{sec-siblinglookup}Modifying the update algorithm for this variant of suffix links, we obtain an
\emph{edge-oriented top-down} (\eotd) variant. The update algorithm is
analogous to the original, except that edge suffix links are set and followed
rather than node suffix links, and the first branch operation of each rescan
avoided as a result. The following points deserve special attention:
\begin{itemize}
\item When an edge is split, the top part should remain the destination of
  incoming suffix links, i.e., the new edge is the bottom part.
\item After splitting one or more edges in an update, finding the correct
  destination for the suffix link of the last new edge (the bottom part of the
  last edge split) requires a \emph{sibling lookup} branch operation, not
  necessary in \notd.
\item Following a suffix link from the endpoint of an edge occasionally
  requires one or more extra rescan operation, in relation to following the
  node-oriented suffix link of the endpoint.
\end{itemize}

The first point raises some implementation issues. Efficient representations
(see e.g.\@ Kurtz's~\cite{kurtzsuftree}) do not implement nodes and edges as
separate records in memory. Instead, they use a single record for a node and
its incoming edge. Not only does this reduce the memory overhead, it cuts down
the number of pointers followed on traversing a path roughly by half. The
effect of our splitting rule is that while the top part of the split edge
should retain incoming suffix links, the new record, tied to the bottom part
should inherit the children. We solve this by adding a level of indirection,
allowing all children to be moved in a single assignment. In some settings
(e.g., if parent pointers are needed), this induces a one
pointer per node overhead, but it also has two potential efficiency
benefits. First, new node/edge pairs become siblings, which makes for a natural
memory-locality among siblings (cf.\@ \emph{child inlining} in
section~\ref{sec-inlining}).\label{sec-notechildcache} Second, the original bottom node stays where it was
in the branching data structure, saving one replace operation. These properties
are important for the efficiency of the \eotd representation.

The latter two points go against the reduction of branch operations that
motivated edge-oriented suffix links, but does not cancel it out. (Cf.\@
table~\ref{tab-opcounts}.)

These assertions are supported by experimental data in
section~\ref{sec-experiments}.

Furthermore, \eotd retains the $O(N)$ total construction time of \notd. To see this, note
first that the modification to edge-oriented suffix links clearly adds at most
constant-time operation to each operation, except possibly with regards to the
extra rescan operations after following a suffix link from the endpoint of an
edge. But Ukkonen's proof of total $O(N)$ rescan time still applies: Consider
the string $t_j\cdots t_i$, whose end corresponds to the active point, and
whose beginning is the beginning of the currently scanned edge. Each downward
move in rescanning deletes a nonempty string from the left of this string, and
characters are only added to the right as $i$ is incremented, once for each
online suffix tree update. Hence the number of downward moves are bounded by
$N$, the total number of characters added.

\newcommand\stylescale{.635}

\begin{figure}[t]
  \begin{center}
\begin{tabular}{@{\hspace{0pt}}c@{\hspace{0pt}}c@{\hspace{0pt}}c@{\hspace{0pt}}c@{\hspace{0pt}}}
\includegraphics[scale=\stylescale]{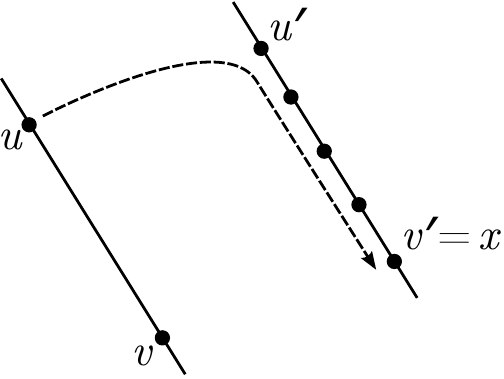}&
\includegraphics[scale=\stylescale]{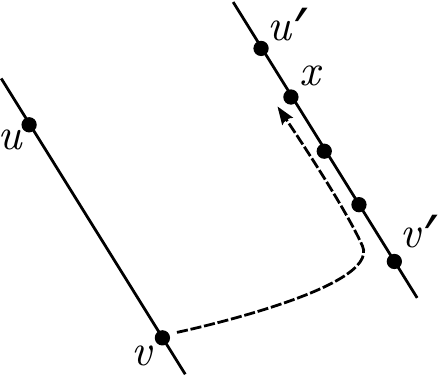}&
\includegraphics[scale=\stylescale]{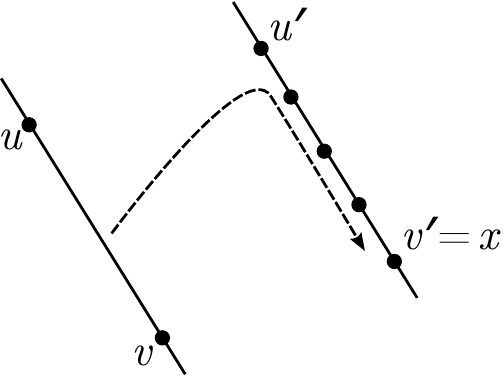}&
\includegraphics[scale=\stylescale]{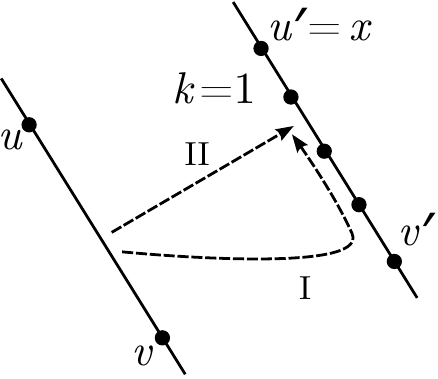}\\[1mm]
\notd & \nobu & \eotd & \eov\\
\end{tabular}

\end{center}
\caption{\label{fig-linkstyles} Examples of moving the active point across a suffix link in
 four schemes, where in each case $x$ is a node that
  coincides with the active point after the move. For \eov, we see two cases,
  before (\textsc{i}) and after (\textsc{ii}) the destination of the suffix
  link is moved  }
\end{figure}

\subsection{Edge-Oriented Variable}\label{sec-eov}

Let $e$ be an edge from node $u$ to $v$, and let $v'$ and $u'$ be nodes
such that node suffix links would point from $u$ to $u'$ and from $v$ to
$v'$. If the path from $u'$ to $v'$ is longer than one edge, the \eotd suffix
link from $e$ would point to the first one, an outgoing edge of
$u'$. Another edge-oriented
approach, more closely resembling \nobu, would be to let $e$'s suffix link to
point to the \emph{last} edge on the path, the incoming edge of $v'$, and use
climb rather than rescan for locating the right edge after following a suffix
link. But this approach does not promise any performance gain over
\nobu.

An approach worth investigating, however, is to allow some freedom in where to
on the path between $u'$ and $v'$ to point $e$'s suffix link. We refer to the
path from $u'$ to $v'$ as the \emph{destination path} of $e$'s suffix link. Given that an
edge maintains the length of the longest string it represents (which is a
normal edge representation in any case), we can use climb or rescan as
required.

We suggest the following \emph{edge-oriented
  variable} (\eov) scheme:
\begin{itemize}
\item When an edge is split, let the bottom part remain the destination of
  incoming suffix links, i.e., let the top part be the new edge. (The
  opposite of the \eotd splitting rule.) This sets a suffix link to the last
  edge on its destination path, possibly requiring
  climb operations after the link is followed.
\item When a suffix link is followed and $c$ edges on its destination path
  climbed, if $c>k$ for a constant $k$, move the suffix link $c-k$ edges up.
\end{itemize}

Intuitively, this approach looks promising, in that it avoids spending time on
adjusting suffix links that are never used, while eliminating the
$\Omega(N^{1.5})$ degeneration case demonstrated for
\nobu~\cite{SenftBranching}. Any node further than $k$ edges away from the top
of the destination path is followed only once per suffix link, and hence the
same destination path can only be climbed multiple times when multiple suffix
links point to the same path, and each corresponds to a separate occurrence of
the string corresponding to the climbed edge labels. We conjecture that the
amortized number of climbs per edge is thus~$O(1)$. However, our experimental
evaluation indicates that the typical savings gained by the \eov approach are
relatively small, and are surpassed by careful application of \eotd.

\section{Branching Data Structure}\label{sec-hashvsll}

Branching efficiency depends on the input alphabet size.  Ukkonen proves $O(N)$ time complexity only under the
assumption that characters are drawn from an alphabet $\alphabet$ where
$|\alphabet|$ is $O(1)$. If $|\alphabet|$ is not constant, \emph{expected} linear
time can be achieved by hashing, as suggested in McCreight's seminal
work~\cite{McR}, and more recent dictionary data
structures~\cite{arbitman2010backyard,hagerup2001deterministic} can be applied
for bounds very close to deterministic linear time. Recursive suffix tree
construction, originally presented by Farach~\cite{FarFOCS} achieves the
same asymptotic time bound as character sorting, but does not support online
construction.

\begin{figure}[t]
  \begin{center}
\includegraphics[width=\textwidth]{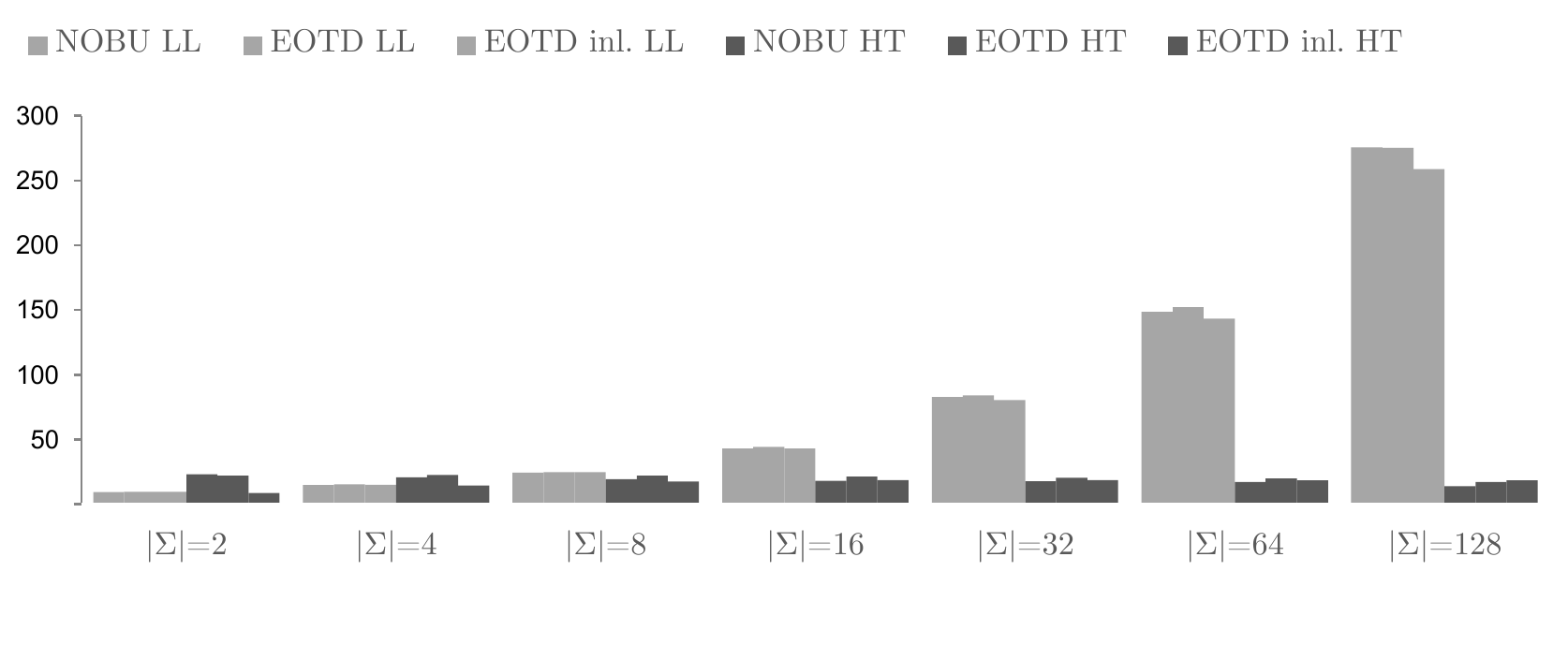}
\end{center}
\vspace{-10mm}
\caption{\label{fig-hashvsll} Comparing linked list (\lili) to hash table
  implementations (\hata) for random files
  different size alphabets. Each file is 50~million characters long, and
  the vertical axis shows runtime in seconds.}
\end{figure}

We limit our treatment to simple schemes based on  linked lists or
hashing since, to our knowledge, asymptotically stronger results have not been shown to
yield a practical improvement. Kurtz~\cite{kurtzsuftree}
observed in 1999 that linked lists appear faster for practical inputs when
$|\alphabet|\leq 100$ and $N\leq 150\,000$. For a lower bound estimate of the
alphabet size breaking point, we tested suffix tree construction on random
strings of different size alphabets. We used a hash table of size $3N$ with linear probing
for collision resolution, which resulted
in an average of less than two hash table probes per insert or lookup
across all files. The results, shown in
table~\ref{fig-hashvsll}, indicate that hashing can outperform linked lists for
alphabet sizes at least as low as 16, and our experiments did indeed show hashing to be
advantageous for the \emph{protein} file, with this size of alphabet. However,
for many practical input types that produce a much lower average suffix tree
node out-degree, the breaking point would be at a much larger $|\alphabet|$.

\label{sec-inlining}

\paragraph*{Child Inlining}

An internal node has, by definition, at least two children. Naturally occurring
data is typically repetitive, causing some children to be accessed more
frequently than others. (This is the basis of the \textsc{ppm} compression
method, which has a direct connection to the suffix tree~\cite{SufComp}.) By
a simple probabilistic argument, children with a high traversal probability
also have a high probability of being encountered first. Hence, we
obtain a possible efficiency gain by storing the first two children of each
node, those that cause the node to be created, as \emph{inline} fields of the
node record together with their first character, instead of including them in
the overall child retrieval data structure. The effect should be particularly strong for \eotd, which, as
noted in section~\ref{sec-eotd}, eliminates the \emph{replace child}
operation that otherwise occurs when an edge is split, and the record
of the original child hence remains a child forever.  Furthermore, if nodes are
laid out in memory in creation order, \eotd's
consecutive creation of the first two children can
produce an additional caching advantage. Note that inline space use is compensated by space
savings in the non-inlined child-retrieval data structure.

When linked lists are used for branching, we can achieve an effect similar to inlining by always inserting new children at the back of the list. This change
has no significant cost, since an addition is made only after an unsuccessful
list scan.

\section{Performance Evaluation on Input Corpora}\label{sec-experiments}

Our target is to keep the number of branch operations low, and
their cost low through lookup data structures with low overhead and good cache
utilization. The overall goal is reducing construction time. Hence, we evaluate
these factors.

\subsection{Models, Measures, and Data}\label{sec-expmodel}

Practical runtime measurement is, on the one
hand, clearly directly relevant for evaluating algorithm behavior. On the other
hand, there is a risk of exaggerated effects dependent on specific hardware
characteristics, resulting in limited relevance for future hardware
development. Hence, we are reluctant to use execution time as the sole performance measure. Another important measure, less dependent on conditions at the
time of measuring, is memory probing during execution. Given the
central role of main memory access and caching in modern architectures, we
expect this to be directly relevant to the
runtime, and include several measures to capture it in our evaluation.


We measure level~3 cache misses using the \emph{Perf} performance counter
subsystem in Linux~\cite{PerfSystem}, which reports hardware events using the
performance monitoring unit of the \cpu. Clearly, with this hardware measure,
we are again at the mercy of hardware characteristics, not necessarily relevant on
a universal scale. Measuring cache misses in a
theoretically justified model such as the \emph{ideal-cache
  model}~\cite{FrigoCacheObliv} would be attractive, but such a model does not easily lend itself
to experiments. Attempts of measuring emulated cache performance using a
virtual machine (\!\emph{Valgrind}) produced spurious results, and the overhead
made large-scale experiments infeasible. Instead, we concocted two simple cache
models to evaluate the locality of memory access: one minimal
cache of only ten 64~byte cache lines with a \emph{least recently used}
replacement scheme (intended as a baseline for the level~one cache of any
reasonable \cpu), and one with a larger amount of cache lines with a simplistic
direct mapping without usage estimation (providing a baseline expected to be at
least matched by any practical hardware).

We measure runtimes of Java implementations kept as similar as possible in
regards to other aspects than the techniques tested, with the 1.6.0\_27
Open\,\textsc{jdk} runtime,  a common contemporary software environment. With current
\emph{hotspot} code generation, we achieve performance on par with compiled
languages by freeing critical code sections of language constructs that allocate
objects (which would trigger garbage collection) or produce unnecessary pointer
dereference. We repeat critical sections ten times per test run, to
even out fluctuation in time and caching. Experiment hardware was a Xeon
E3-1230 v2 3.3GHz quadcore with 32\,kB per core for each of data and
instructions level~1 cache, 256\,kB level~2 cache per core, 8\,MB shared level~3
cache, and 16\,GB 1600\,MHz \textsc{ddr}3 memory. Note that this configuration
influences only runtime and physical cache (table~\ref{tab-runtimes} and the
first two bars in each group of fig.~\ref{fig-maidiagram}); other measures are
system independent.

We evaluate over a variety of data in common use for testing string processing
performance, from the \emph{Pizza \& Chili}~\cite{PizzaChiliCorpus} and
\emph{lightweight}~\cite{LightweightCorpus} corpora. In order to evaluate a
degenerate case for \nobu, we also include an adversary input constructed for
the pattern $T=ab^{m^2}abab^2ab^3\cdots ab^ma$ (with $m=4082$ for a 25~million
character file), which has $\Omega(N^{1.5})$ performance in this
scheme~\cite{SenftBranching}.

\subsection{Results}\label{sec-results}

\begin{figure}[t]
  \begin{center}
\includegraphics[width=\textwidth]{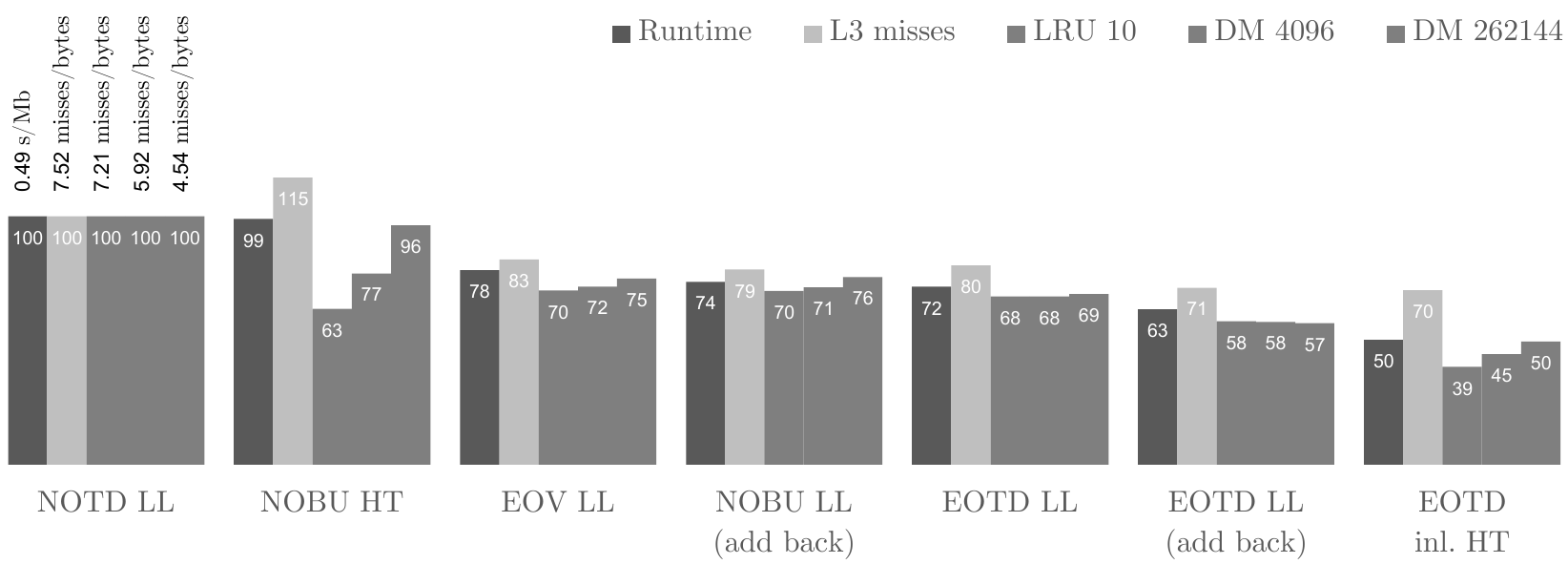}
\end{center}
\caption{\label{fig-maidiagram} Main comparison diagram for performance
  measures across the file set, excluding \emph{adversary}.  Branch
  data structures are either linked lists (\lili, where new entries are added
  at the front unless \emph{add back} is stated), or a single hash table.}
\vspace{1mm}
\end{figure}

\begin{table}
  \begin{center}
    \scriptsize
    \addtolength{\tabcolsep}{.3ex}
    \begin{tabular}{lrrrrrrr}
	&	$\text{size}\cdot$&	\notd&	\eotd &	\eov  &	\eov &	\nobu & move down\\
	File&	$10^{-6}$& rs&	rs+sl&	rs&	climb&	climb&	branch ops \\[.5ex] \hline
		chr22$^{\text{B1}}$&	34.55&	29\,569\,178&	18\,927\,812&
                318\,499&	33\,064\,133&	33\,669\,019&	35\,053\,371 \tstrut \\
		dna$^{\text{A1}}$&		104.86&	87\,681\,116&	58\,585\,203&	236\,634&	100\,172\,270&	100\,736\,053&	111\,372\,537 \\
		dblp$^{\text{A2}}$&		104.86&	54\,757\,925&	14\,743\,305&	32\,980&	55\,418\,573&	55\,594\,399&	73\,784\,654 \\
		rctail96$^{\text{B2}}$&	114.71&	74\,993\,651&	20\,777\,946&	86\,190&	71\,863\,312&	72\,128\,546&	70\,211\,436 \\
		jdk13c$^{\text{B3}}$&	69.73&	50\,659\,938&	6\,678\,647&	54\,174&	49\,300\,828&	49\,413\,385&	28\,044\,490 \\
		sources$^{\text{A3}}$&	104.86&	80\,270\,528&	30\,753\,392&	191\,755&	75\,419\,031&	75\,953\,764&	70\,537\,447 \\
		w3c2$^{\text{B3}}$&		104.20&	80\,056\,887&	12\,933\,438&	57\,108&	75\,773\,161&	75\,904\,742&	41\,111\,077 \\
		english$^{\text{A4}}$&	104.86&	86\,528\,338&	43\,803\,204&	109\,151&	78\,451\,578&	78\,998\,269&	85\,577\,767 \\
		etext$^{\text{B4}}$&	105.28&	73\,446\,539&	40\,782\,335&	106\,811&	73\,482\,182&	74\,097\,636&	99\,131\,563 \\
		howto$^{\text{B4}}$&	39.42&	28\,590\,381&	13\,523\,660&	89\,650&	27\,703\,460&	27\,944\,722&	32\,676\,237 \\
		rfc$^{\text{B4}}$&		116.42&	88\,716\,588&	32\,739\,584&	452\,280&	83\,618\,480&	84\,486\,767&	77\,334\,572 \\
		pitches$^{\text{A5}}$&	55.83&	47\,744\,716&	21\,303\,582&	279\,505&	42\,615\,067&	43\,081\,419&	46\,777\,918 \\
		proteins$^{\text{A6}}$&	104.86&	74\,912\,821&	39\,662\,469&	31\,075&	70\,942\,644&	71\,016\,405&	111\,979\,688 \\
		sprot34$^{\text{B7}}$&	109.62&	70\,190\,029&	20\,737\,274&	45\,034&	69\,274\,605&	69\,425\,197&	78\,927\,702 \\
		adversary$^{\text{A8}}$&	25.00&	41\,662\,928&	16\,323&	8\,313\,003&	41\,654\,774&	68033\,898\,010&	12\,249 \\	
      \end{tabular}
      \end{center}

      \caption{\label{tab-opcounts} Operation counts.  \emph{rs}: rescan branch
        operations, \emph{sl}: extra
        sibling lookup (see section~\ref{sec-siblinglookup}), \emph{move down}:
        branch operations outside of rescan. Files from the
        \emph{Pizza and Chili Corpus} ($^{\text{A}}$) and \emph{Lightweight
          Corpus} ($^{\text{B}}$). File categories are \textsc{dna}~($^{\text{1}}$),
        \textsc{xml}~($^{\text{2}}$), source code~($^{\text{3}}$), text~($^{\text{4}}$),
        \textsc{midi}~($^{\text{5}}$), proteins~($^{\text{6}}$),
        database~($^{\text{7}}$), and \nobu
        adversary~($^{\text{8}}$).}
\end{table}

Fig.~\ref{fig-maidiagram} shows performance across seven implementations and
five performance measures (explained in section~\ref{sec-expmodel}), which we
deem to be relevant for comparison. It summarizes the runtimes (also in
table~\ref{tab-runtimes}) and memory access measures by taking averages across
all files except \emph{adversary}, with equal weight per file. The bars are
scaled to show percentages of the measures for the basic \notd implementation,
which is used as the benchmark. The order of the implementations when ranked by
performance is fairly consistent across the different measures, with some
deviation in particular for the hardware cache measure and smaller-cache
models. The hardware cache measurement comes out as a relatively poor predictor
of performance; by the numbers reported by Perf, the hardware cache even appears
to be outperformed by our simplistic theoretical cache model.

We detect only a minor improvement of \eotd \lili implementations in relation to
\nobu \lili, while inline \eotd \hata provides a more significant
improvement. Note, however that for \nobu, the \hata implementation is much
worse than the \lili implementation, while the reverse is true for \eotd. This
can be attributed to the different hash table use and the particular
significance of inlining, noted in section~\ref{sec-inlining}. The fact that
\eotd \hata without inlining (not in the diagram) is not clearly better than
\nobu \hata stands to confirm this. Although table~\ref{tab-runtimes} shows that
\eotd \lili beats its \hata counterpart for files producing a low average
out-degree in $\ST$ (because of a small alphabet and/or high repetitiveness), the
robustness of hashing (cf.\@ fig~\ref{fig-hashvsll}) has the greater impact on
average. We have included results to show the impact of the \emph{add to back}
heuristic in \eotd \lili, which also produced a slight improvement for \nobu (not
shown in diagram), as expected.

The operation counts shown in table~\ref{tab-opcounts} generally confirm our
expectations.  (Branch counts include moves down from $\topnode$ to the
root, in order to match Senft and Dvořák's corresponding
counts~\cite{SenftBranching}.) \eov yields a large rescan reduction, even for the adversary
file, which makes it an attractive alternative to \nobu when branching is very
expensive.  We found the exact choice of the $k$ parameter of \eov not to be
overly delicate. All values shown were obtained with $k=5$.

\section{Conclusion}\label{sec-concl}

It is possible to significantly improve online suffix tree construction time  
through modifications that target reducing branch operations and cache  
utilization, while maintaining linear worst-case time complexity. In many  
applications, our representation variants should be directly applicable for
runtime reduction. Interesting topics remaining to explore are how our
techniques for, e.g., suffix link orientation, fit into the compromise game of
time versus space in succinct representations such as compressed suffix trees,
and comparison to off-line construction.

\begin{table}[t]
\begin{center}
    \scriptsize
     \addtolength{\tabcolsep}{.3ex}
    \begin{tabular}{lrrrrrrrrrrr}
		&             \notd	&	\notd&	\nobu&	\nobu&	\nobu&
                \eov  &	\eov &	\eotd&	\eotd &	\eotd&	\eotd\\
	File&	     &	\hata&	             &    back & \hata  &	\lili &	        \hata &	\lili&	         \hata&	 back&	     inl. \hata \\ \hline
	chr22&	11.43&	16.73&	8.66& 9.00 &	13.72&	9.08&	14.26&	8.96&	14.40&	8.80&	8.91 \\
	dblp&	29.31&	35.56&	22.41& 21.90 &	30.60&	23.60&	32.15&	20.35&	26.55&	17.67&	16.91 \\
	dna&	40.37&	60.76&	30.65& 32.12 &	51.66&	32.70&	53.77&	31.60&	53.37&	30.97&	32.89 \\
	english&	64.26&	50.99&	45.65& 46.36 &	42.11&	47.47&	43.34&	42.70&	42.77&	36.64&	26.21 \\
	etext&	64.96&	50.15&	47.68& 46.44 &	42.43&	50.37&	44.30&	45.56&	43.38&	39.06&	27.67 \\
	howto&	21.74&	15.43&	16.12& 15.09 &	12.64&	16.48&	12.92&	15.33&	12.50&	12.56&	7.61 \\
	jdk13c&	7.97&	23.24&	6.39& 6.53 &	19.29&	6.97&	20.24&	5.72&	14.46&	5.27&	6.76 \\
	pitches&	46.65&	21.34&	34.66& 28.98 &	18.40&	35.38&	19.07&	34.08&	17.26&	26.34&	10.55 \\
	proteins&	104.49&	49.60&	74.30& 74.46 &	41.95&	76.49&	44.27&	75.73&	46.18&	70.55&	31.97 \\
	rctail96&	35.67&	44.76&	26.72& 26.54 &	37.44&	27.61&	38.35&	24.59&	31.32&	21.18&	18.35 \\
	rfc&	52.58&	50.81&	38.78& 37.22 &	42.96&	40.55&	44.14&	37.18&	39.99&	29.45&	21.82 \\
	sources&	44.21&	44.23&	32.71& 30.12 &	37.24&	34.27&	38.63&	31.49&	34.28&	24.70&	17.76 \\
	sprot34&	50.19&	42.65&	38.40& 37.92 &	37.37&	39.82&	38.50&	36.71&	33.66&	33.24&	21.24 \\
	w3c2&	18.98&	39.84&	14.38& 15.19 &	33.13&	14.89&	33.73&	12.91&	24.98&	11.41&	10.47 \\
	adversary&	1.30&	7.89&	267.50&	266.16&	296.42&	1.64&	8.07&	1.40&	5.10&	1.39&	1.34 \\
    \end{tabular}
    \end{center}
    \caption{\label{tab-runtimes}Running times in seconds for the same files as
    table~\ref{tab-opcounts}}
      \vspace{-2mm}
\end{table}

\end{document}